\documentclass[aps,prl,twocolumn,groupedaddress,floatfix,showpacs]{revtex4}
\usepackage{graphics,graphicx}
\usepackage{color}
\usepackage{amssymb,amsmath}
\usepackage{bm}
\newcommand{\req}[1]{Eq.~(\ref{#1})}

\newcommand{\vare}{\varepsilon}
\newcommand{\cF}{{\cal F}}
\newcommand{\Del}{{\mbox{\footnotesize $\Delta$}}}

\newcommand{\Dc}{{\Del C/T_c}}

\newcommand{\be}{\begin{equation}}
\newcommand{\ee}{\end{equation}}
\newcommand{\bea}{\begin{eqnarray}}
\newcommand{\eea}{\end{eqnarray}}
\begin{document}
\unitlength = 1mm
\title{Specific heat jump at superconducting transition in the presence of Spin-Density-Wave in iron-pnictides}
\author{M.~G.~Vavilov$^1$, A.~V.~Chubukov$^1$, and  A.~B.~Vorontsov$^2$ }
\affiliation{$^1$~Department of Physics,
             University of Wisconsin, Madison, Wisconsin 53706, USA\\
$^2$~Department of Physics, Montana State University, Bozeman, MT, 59717, USA}

\date{April 26, 2011}

\begin{abstract} 
We analyze the magnitude of the specific heat jump $\Del C$  at the
superconducting transition temperature $T_c$ in the situation when
superconductivity develops in the pre-existing antiferromagnetic phase. We show
that  $\Del C/T_c$ differs from the BCS value and is peaked at the tri-critical
point where this coexistence phase first emerges.  Deeper in the magnetic
phase, the onset of coexistence, $T_c$, drops and $\Del C/T_c$ decreases,
roughly as $\Del C/T_c \propto T^2_c$ at intermediate $T_c$ and exponentially
at the lowest $T_c$, in agreement with the observed behavior of $\Del C/T_c$ in
iron-based superconductors.  
\end{abstract}
\maketitle

\emph{Introduction}.
The magnitude and the doping dependence of  the specific heat jump at
the superconducting transition temperature $T_c$ is one of unexplained
phenomena in novel iron-based superconductors (FeSCs)~\cite{johnson}. In BCS
theory $\Dc \simeq 1.43 \gamma$,  
where $\gamma = \pi^2 N_F/3$ is the
Sommerfeld coefficient, and $N_F$ is the total quasiparticle density of states
(DoS) at the Fermi surface (FS) in the normal state.   
Although the behavior of FeSCs is in
many respects consistent with BCS theory, the experimental values of $\Dc$
vary widely between different compounds, ranging between  
$1~ mJ/(mol\cdot K^2)$ in underdoped $Ba(Fe_{1-x}Ni_x)_2As_2$~\cite{budko_09}
and $100~ mJ/(mol\cdot K^2)$ in optimally hole-doped $Ba_{1-x}K_xFe_2As_2$~\cite{large}. 
Such huge variations may be partly  due to differences in
$\gamma$, which were indeed reported to be larger in hole-doped FeSCs~\cite{large,hardy_10}.  
Yet, even for a given material, e.g.,  $Ba(Fe_{1-x}Ni_x)_2As_2$ or
$Ba(Fe_{1-x}Co_x)_2As_2$  the magnitude of $\Dc$ 
 is peaked near optimal doping $x_{opt}$ and rapidly decreases,
approximately as $\Dc \propto T^2_c$,  at smaller and larger
dopings~\cite{budko_09}.

This rapid 
and non-monotonic 
variation of $\Dc$ over a relatively small range of $0.03< x< 0.12$ 
is unlikely to 
be attributed to change in $\gamma$ and has to be explained by other effects. 
According to one proposal~\cite{kogan}, $\Dc \propto T^2_c$ is caused by  
interband scattering off
non-magnetic impurities, which is pair-breaking for  $s^\pm$ pairing. 
However,  $\Dc \propto T^2_c$  only holds 
in the gapless superconductivity regime,\cite{kogan,vvc_rho} when $T_c$ is
already reduced by more than factor of five as compared to $T_c$ for a clean case.  Before that,
the reduction of  $\Dc$ by impurities is rather mild. Experimentally, the
reduction is strong immediately away from optimal doping and, moreover, occurs 
on both sides from the optimal doping. 
 This reduction is difficult to explain by impurity scattering.
 Large value of $\Dc$ at $x_{opt}$ may also be
due to strong coupling effects~\cite{hardy_10}. This is certainly a possibility, but we note
that $\Dc$ is a non-monotonic function of the coupling~\cite{marsiglio}
and at strong coupling  is actually  smaller than the BCS value. 

\begin{figure}
\centerline{\includegraphics[width=0.95\linewidth]{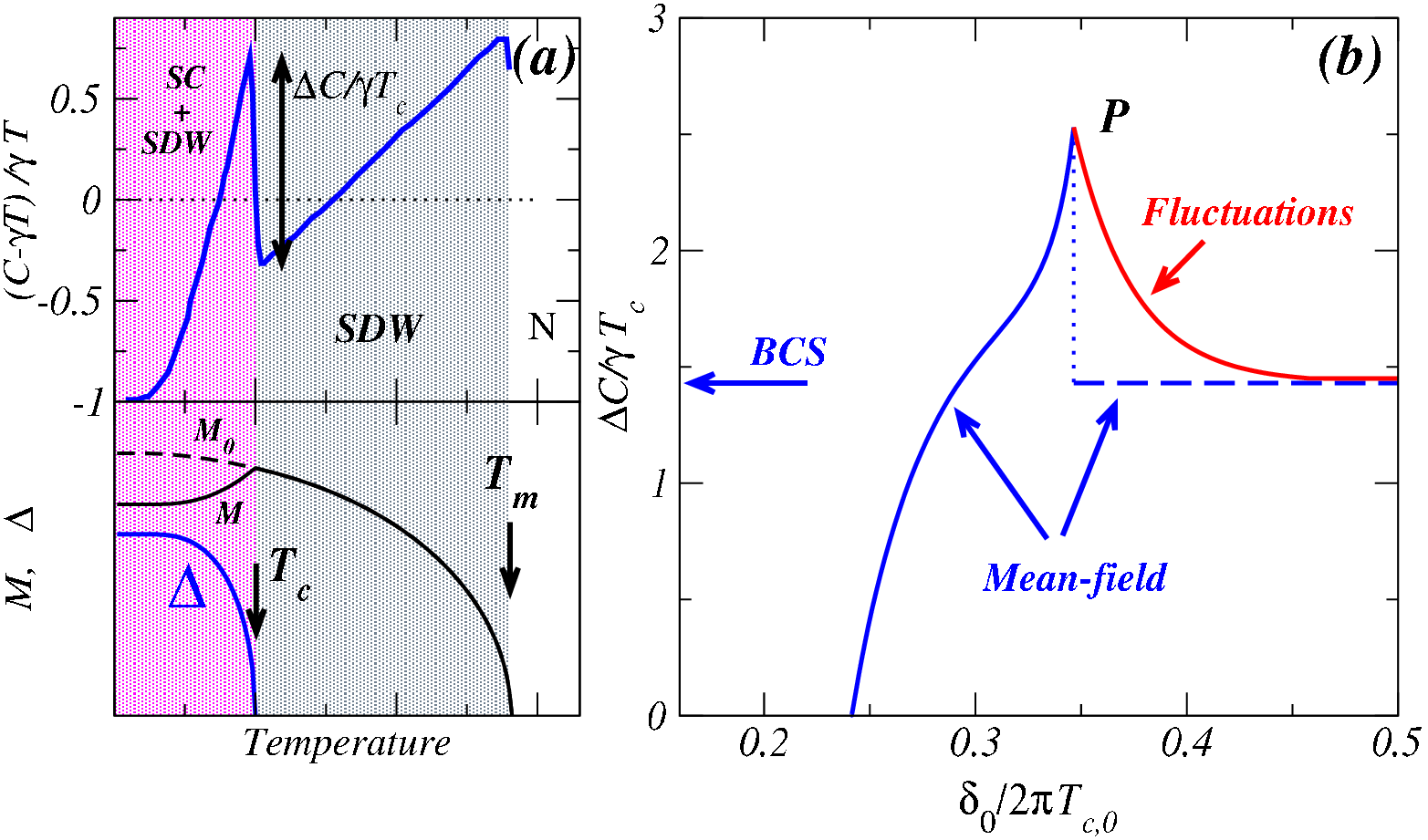}}
\caption{(Color online) 
(a) The specific heat $C(T)$ 
 and SC  and SDW order parameters $\Delta$ and $M$
as functions of $T$. 
We 
consider 
the jump of $C(T)$ 
at the onset of SC. 
(b) The behavior of $\Del C/(\gamma T_c)$ as a function of 
$\delta_0$, which scales with doping.  
In a mean-field theory, $\Dc$ is discontinuous at the end point of the coexistence state (P) 
and jumps back to BCS value $\Dc \simeq 1.43\gamma$ at larger dopings
(dashed horizontal line). Beyond  mean-field, 
paramagnetic fluctuations smear the discontinuity of $\Dc$ and transform it
into a maximum, as schematically shown by the solid line.}
\label{fig:1}
\end{figure}

\begin{figure}
\centerline{\includegraphics[width=0.90\linewidth]{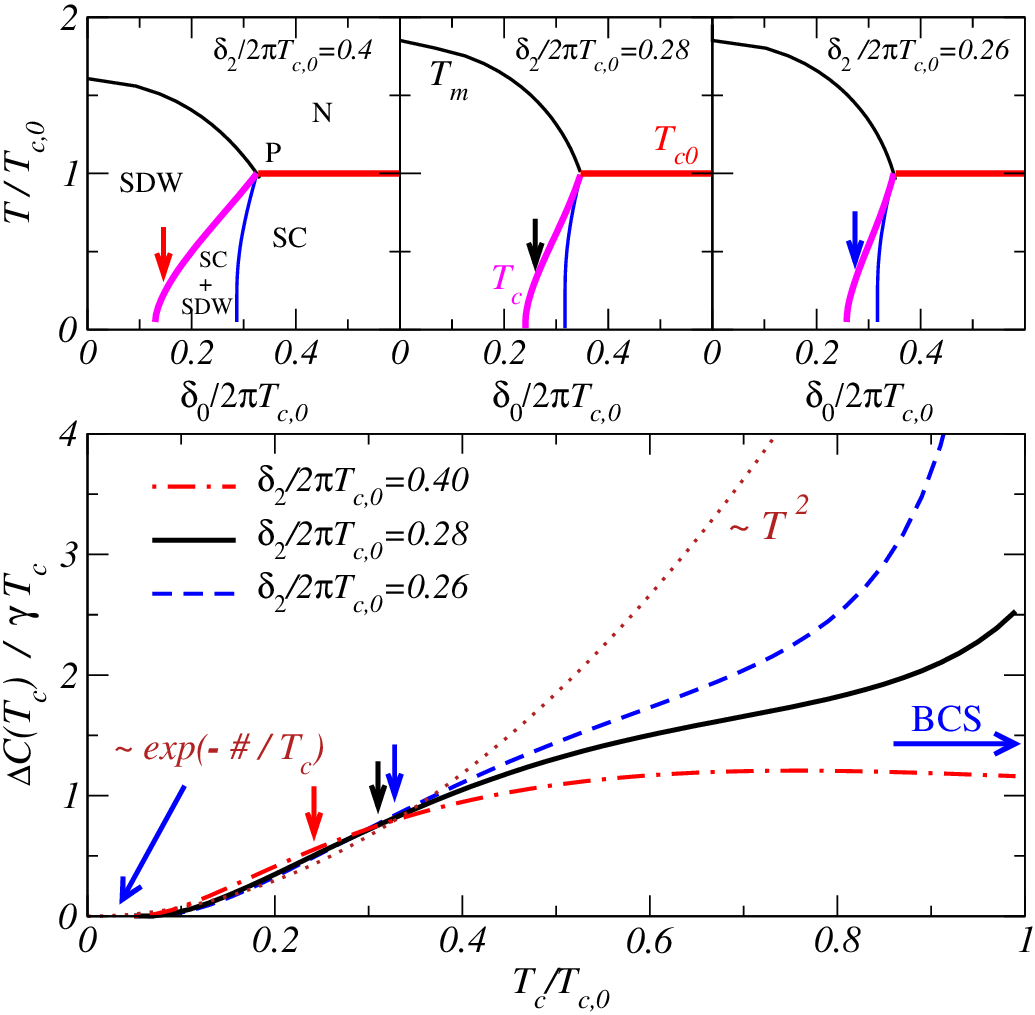}}
\caption{(Color online) \emph{Top:}  
The phase diagram in $T$-$\delta_0$ plane for
$T_{m,0}/T_{c,0}=2$ 
and several $\delta_2/(2\pi T_{c,0})=0.4,\ 0.28,\ 0.26$,
corresponding to wide, medium, and narrow doping ranges of the coexistence
phase. SDW, SC and the 
SDW+SC 
phases meet at the tri-critical point $P$ 
(in this case also meeting normal (N) state). 
\emph{Bottom:} The behavior of $\Delta C/\gamma T_c$
vs $T_c/T_{c,0}$ in the coexistence region for these $\delta_2$.
The arrows indicate $T_c$, below which the whole FS is gapped by SDW.
As $T_c$ is lowered 
through 
this value, $\Dc$ decreases,  as $T^2_c$ 
at intermediate $T_c$ and exponentially at lower $T$. 
Near the tri-critical point,  
$\Dc$ may well exceed  the BCS value $1.43\gamma$.
}
\label{fig:2}
\end{figure}

We propose a different explanation. We  argue that the origin of strong doping
dependence of $\Dc$  is the coexistence of spin-density-wave (SDW) magnetism
and $s^\pm$ superconductivity (SC).~\cite{zlatko,fernandes,anton} In
$Ba(Fe_{1-x}Ni_x)_2As_2$, $Ba(Fe_{1-x}Co_x)_2As_2$ and, possibly, in other
FeSCs, optimal doping $x_{opt}$ nearly coincides with the end point of the
coexistence region (tri-critical point).~\cite{fernandes} We analyzed the
behavior of  $\Dc$ near $x_{opt}$ within a mean-field BCS-like theory and found
that $\Dc$ is by itself discontinuous and jumps by a finite amount  when the
system enters the coexistence region, see Figs.~\ref{fig:1} and \ref{fig:2}. 
The magnitude of the jump depends on microscopic parameters of the system and for a wide range of parameters, $\Dc$ 
exceeds the BCS value.
Beyond a mean-field
treatment, paramagnetic fluctuations transform the discontinuity in $\Dc$ at
$x_{opt}$ into a maximum, such that $\Dc$ smoothly 
decreases on both sides of optimal
doping $x_{opt}$, as illustrated in Fig.~\ref{fig:1}.

We also examined the behavior of $\Dc$ along 
the entire transition into the coexistence state.  
We found that $\Dc$ decreases together with $T_c$,  and eventually becomes smaller that its BCS value.
Deeper in the SDW region, $\Dc \propto T^2_c$ and  vanishes exponentially at the 
lowest $T_c$, see Fig.~\ref{fig:2}.  
The explanation of this behavior goes beyond a standard paradigm that $T_c$ and $\Delta C$ decrease because FS  available for SC is modified by SDW.
If that was the only effect, then the DoS would not change significantly and $\Del C/T_c$ would only weakly depend on $T_c$.  
We found, however, that the SC transition line 
{\it necessarily} 
enters the region in which  SDW order gaps out the whole FS.
In this region, SC appears 
 because the system finds that it is energetically advantageous to 
decrease the magnitude of the SDW order parameter
 and develop a non-zero SC order parameter, even in the absence of the FS.
Because all low-energy states in this region
are gapped at $T_c$, the thermodynamic characteristics, including $\Dc$, are exponentially small. 
We found that the precursors of this behavior develop at higher  $T_c$, when the reconstructed FS is still present. As a result, $\Dc$ decreases as the SC temperature, $T_c$, drops. 

The behavior of $\Dc$ outside the coexistence region is likely to be 
a combination of several effects.  When paramagnetic fluctuations 
weaken, $\Dc$ reduces to its BCS value.
Further decrease of $\Dc$ is partly due to impurities,\cite{kogan,vvc_rho} 
and partly due to shrinking of the hole FSs and to the fact
that at larger $x$ the gap along electron FSs 
 becomes more anisotropic.

\emph{The method}. To obtain $\Delta C$, we expand 
 the free energy in powers of the SC order parameter $\Delta$  to order $\Delta^4$. When
 the SC transition occurs from a pre-existing SDW state, the expansion reads  
\be 
\frac{{\cal F}(\Delta,M_0)}{N_F}  =\frac{{\cal F}_0}{N_F}  + \alpha_\Delta (M_0,T) \Delta^2+\eta(M_0,T) \Delta^4,
\label{eq:GLinDelta}
\ee
where ${\cal F}_0 ={\cal F} (0,M_0)$ is the free energy of a pre-existing
 SDW state,
$M_0=M_0(T)$ is the SDW order parameter which minimizes 
${\cal F}(0,M)$, 
 and $\eta$ includes 
 the feedback of the finite SC order parameter on the SDW state, 
$M^2 = M_0^2 - {\cal{O}}(\Delta^2)$, 
 see Fig.~\ref{fig:1}(a).
 The $T_c$ is given by  $\alpha_\Delta (M_0(T_c),T_c)=0$ and the specific heat jump is 
\be \label{eq:heatjump}
 \frac{\Del C}{T_c}=\frac{3\gamma} {2 \pi^2\eta}\left(\frac{d\alpha_\Delta}{dT}\right)^2_{\alpha_\Delta =0}, \quad
 \frac{d\alpha_\Delta}{dT}= \frac{\partial \alpha_\Delta}{\partial T}+\frac{\partial \alpha_\Delta}{\partial M_0^2}\frac{dM_0^2}{dT}.
\ee
As $\eta\to0$, $\Delta$ 
appears discontinuously 
and the transition between SDW and SC becomes first order. 

To obtain actual expressions for $\alpha_\Delta$ and $\eta$, we need to specify the band structure of a material. 
Since our goal is to demonstrate the discontinuity of $\Dc$ at $x_{opt}$ and
the reduction of $\Dc$ along 
the coexistence onset, 
we adopt a simplified 2D two-band model  with the
hole-like band  near the center of the Brillouin zone (BZ), with $\xi_h = \mu_h - k^2/2m_h$, 
and electron-like band near the corner of the BZ, 
with $\xi_e = -\mu_e + k^2_x/2m_x + k^2_y/2m_y$, here $k_x$ and $k_y$ are deviations from $(\pi,\pi)$.
The same model has been considered in Refs.~\cite{fernandes,anton,parker} on the coexistence of
SDW and SC orders.  At perfect nesting, $\xi_e = - \xi_h$, while 
for a non-perfect nesting $\xi_e = - \xi_h + 2\delta_\varphi$ near the FS, and  
$\delta_\varphi = \delta_0+\delta_2 \cos 2\varphi$
captures the difference in the chemical potentials and 
in electron $m_{x,y}$ and hole $m_h$ masses via $\delta_0$, and anisotropy 
in  $m_{x}$ and $m_y$  via $\delta_2$. Without loss of generality, we
assume that $\delta_0$ changes with doping,
but the ellipticity parameter $\delta_2$ remains intact.
 We 
 use a model with
 four-fermion interactions in SDW
and SC channels~\cite{anton,CEE,Wang2009}, decompose them using SDW and SC order parameters $M$
and $\Delta$, and express couplings in terms of transition temperatures
$T_{c,0}$ to the SC state in the absence of SDW and $T_{m,0}$ to the 
perfectly nested SDW state, $\delta_{0,2}=0$) in the absence of SC. Note  
that the actual $T_m$ decreases, even in the absence of SC, when $\delta_0$ and $\delta_2$ increase, while
 $T_{c,0}$ is the transition temperature at the
tri-critical point indicated by $P$ in Fig.~\ref{fig:2}.
 
The free energy for such a model has the form~\cite{anton}
\begin{equation}
\label{eq:F2band}
\begin{split}
&
\frac{\cF(\Delta, M)}{N_F}=  
\frac{\Delta^2}{2} \ln\frac{T}{T_{c,0}} + \frac{ M^2}{2} \ln\frac{T}{T_{m,0}}
\\
&
-
 2\pi T \sum_{\vare_n>0} 
{\rm Re}\left\langle
\sqrt{(E_n+i\delta_\varphi)^2+M^2} - \vare_n
- \frac{\Delta^2+M^2}{2\vare_n}
\right\rangle , \nonumber
\end{split}
\ee
where $E_n=\sqrt{\vare_n^2+\Delta^2}$, $\vare_n=\pi T(2n+1)$ are the Matsubara
frequencies ($n=0,\pm1,\pm2,\dots$), 
and $\langle\dots\rangle$ denotes averaging over 
$\varphi$ along FSs. 
For this functional we find 
 \begin{equation}
\begin{split}
\label{eq:alphas}
&\alpha_\Delta 
 =  \frac{\partial\cF}{\partial (\Delta^2)}
  =\frac{1}{2}\ln \frac{T}{T_{c,0}}  + 
 \pi T \sum_{\vare_n>0} \frac{1}{\vare_n}\left(1-K\right) \,, 
\\
&\eta (M_0, T) = A-C^2/B,
\end{split}
\end{equation}
 where
 \begin{equation}\label{eq:ABC}
\begin{split}
 K &= \left\langle{\rm Re}\frac{\vare_n+i\delta_\varphi}{\sqrt{(\vare_n+i\delta_\varphi)^2+M_0^2}}\right\rangle,\\
A & = \frac{1}{2}\frac{\partial^2\cF}{\partial(\Delta^2)^2}
 =\sum_{\vare_n>0} \frac{\pi T}{4\vare_n^3}{\rm Re}\left\langle
\frac{(\vare_n+i\delta_\varphi)^3+i\delta_\varphi M^2_0}{((\vare_n+i\delta_\varphi)^2+M_0^2)^{3/2}}\right\rangle,\\
B & = \frac{1}{2}\frac{\partial^2\cF}{\partial( M^2)^2}
 \sum_{\vare_n>0} {\rm Re}\left\langle \frac{\pi T}{4((\vare_n+i\delta_\varphi)^2+M^2_0)^{3/2}}
\right\rangle ,
\\
C & =  \frac{1}{2}\frac{\partial^2\cF}{\partial(\Delta^2)\partial( M^2)} 
 =\sum_{\vare_n>0} {\rm Re}\left\langle \frac{\pi T(\vare_n+i\delta_\varphi)/4\vare_n} {((\vare_n+i\delta_\varphi)^2+M^2_0)^{3/2}}
\right\rangle .
 \end{split}\end{equation}
 The derivatives are taken at $\Delta=0$ and $M=M_0$ with
 $M_0$ defined by
\be 
\label{eq:SCM0}
\ln\frac{T_{m,0}}{T}=2 \pi T\sum_{\vare_n>0}{\rm Re} \left\langle\frac{1}{\vare_n} -\frac{1}{\sqrt{(\vare_n+i\delta_\varphi)^2+M_0^2}}
\right\rangle
.
\ee
In the absence of SDW, $M_0 \equiv 0$, $d\alpha_\Delta/dT=\partial \alpha_\Delta/\partial T$, 
$\eta = A (M_0 =0) = 7 \zeta (3)/(32 \pi^2 T^2)$, 
and we 
 reproduce
 the BCS result $\Dc = 1.43 \gamma$.
 To obtain $\Dc$ 
 inside SDW phase 
  we solve 
Eq.~(\ref{eq:SCM0})  for $M_0^2 (T)$, 
insert the result into Eqs.~(\ref{eq:ABC}),
evaluate 
 $d\alpha_\Delta/dT$ and $\eta$  
 and  finally use  Eq.~(\ref{eq:heatjump}). 
$\Dc$ depends on three input
parameters $\delta_0, \delta_2$, and $T_{m,0}/T_{c,0}$, and generally differs
from the BCS value.

\emph{Results}. 
We present 
 $\Dc$ as function  of $\delta_0$
 for   fixed $\delta_2$ and $T_{m,0}/T_{c,0}$ in Fig.~\ref{fig:1}(b). 
It grows from zero value at the low-temperature onset of the coexistence phase
and reaches its maximum at the tri-critical point, where  SDW order disappears.
At this $\delta_0$, $T_c$ reaches $T_{c,0}$ and $\Dc$  jumps 
 to the BCS value. 

Plotted as a function of $T_c/T_{c,0}$ in Fig.~\ref{fig:2}, 
$\Dc$ shows exponential behavior at small $T_c$ and approximate 
$T_c^2$ behavior at intermediate 
$T_c/T_{c,0} \lesssim 0.5$.  
The magnitude of 
 $\Dc$ at $T_{c,0}$ increases when the width of the
coexistence region shrinks. This can be easily understood, 
since shrinking of
the coexistence region brings the system closer to a first order transition
between SDW and SC 
 at which the entropy itself becomes discontinuous at $T_c$,
and $\Dc$ diverges.   In the opposite limit, when the width of the coexistence
range is the largest, the magnitude of 
 $\Dc$
is smaller and can even be below the BCS value.

The behavior of $\Dc$  at small $T_c$ and at $T_c \approx T_{c,0}$ 
can be understood analytically. 
We first focus on the low $T_c$ limit, $T_c\ll T_{c,0}$. We argue that at low
$T_c$ a conventional reasoning that the mixed SDW/SC state exists because FS
reconstruction associated with SDW order still leaves pieces of the FS
available for SC does not work because the SDW state immediately above $T_c$ is
already fully gapped. Indeed, to a logarithmical accuracy, 
the condition $\alpha_\Delta (M_0,T_c) =0$  becomes
\begin{equation}
\label{eq:alphasLT}
2\pi T_c\sum_{\vare_n>0}\frac{1}{\vare_n} \left(1-{\rm Re}\left\langle\frac{\delta_\varphi}{\sqrt{\delta_\varphi^2-M_0^2}}\right\rangle \right) 
= \ln\frac{T_{c,0}}{T_c} 
\end{equation}
 which is only satisfied if 
$M_0>{\rm max}\{\delta_\varphi\}=\delta_0+\delta_2$,
 \emph{i.e.} the SDW state is fully gapped~\cite{anton}.

The coexistence state emerges from the fully gapped SDW state because, 
when $\eta >0$ and  $\alpha_\Delta>0$, 
it is energetically
advantageous to gradually reduce the magnitude of the SDW order
$M$ below $M_0$,
thus change sign of $\alpha_\Delta$ and
create non-zero SC order $\Delta$. The contribution to
the free energy from SC ordering comes from the rearrangement of quasiparticle
states above the 
gap, and the number of such quasiparticles exponentially  decreases with lowering $T_c$.
As a result, the magnitude of the specific heat discontinuity at $T_c$ 
also becomes exponentially small, 
$\Dc \propto  \exp[-(M_0-\delta_0-\delta_2)/T_c]$, consistent with Fig. \ref{fig:2}. 
We emphasize that as long as SDW--coexistence transition is of second order, $\eta>0$, 
the exponential behavior at low $T_c$ 
always holds. 
The behavior of $\Dc$ at larger $T$ is less universal, but 
the reduction of $\Dc$ always begins even before the condition $M_0 >\delta_0 + \delta_2$ is reached, see  Fig.~\ref{fig:2}.  
For smaller $\eta$, $\Dc$ decreases with $T_c$ over the whole range $T_c < T_{c,0}$.

\begin{figure}
\centerline{\includegraphics[width=0.95\linewidth]{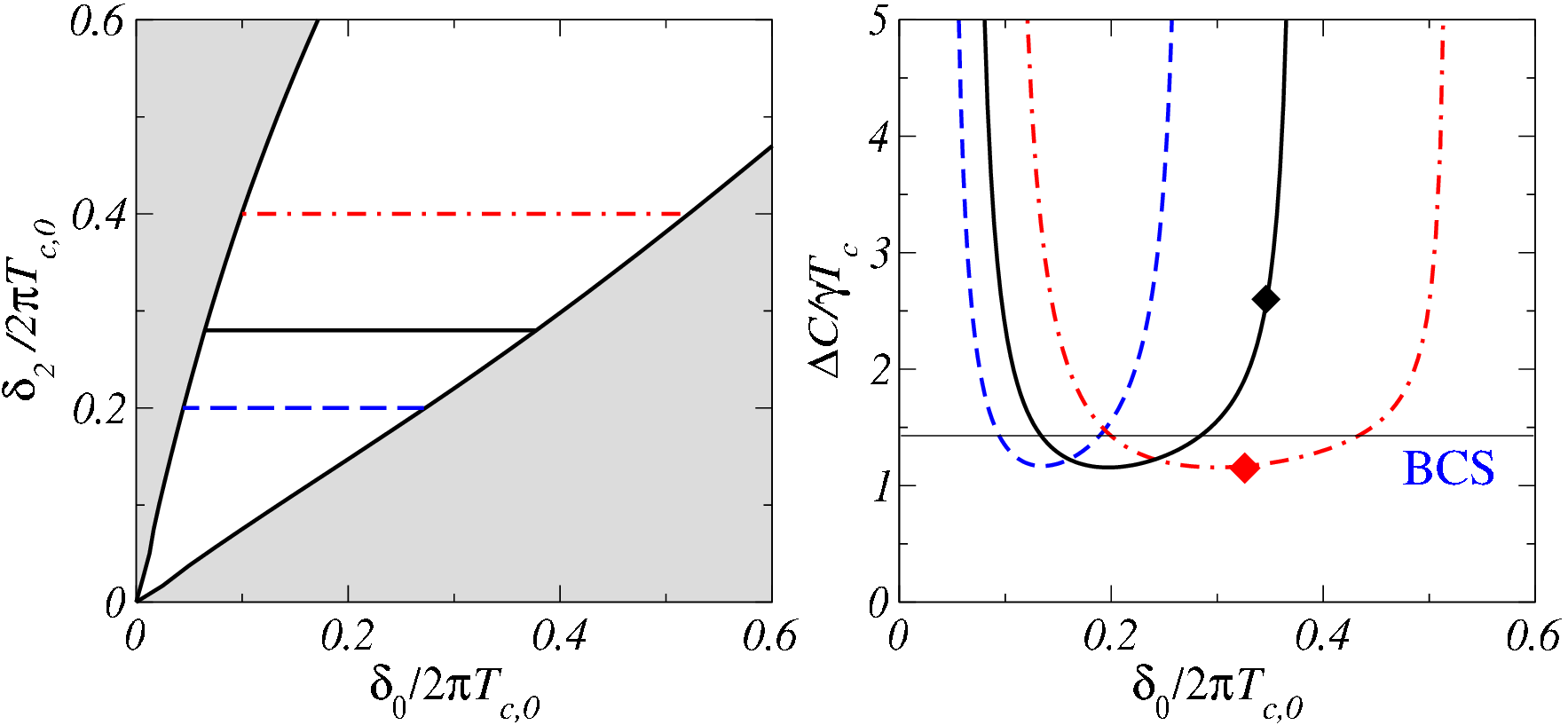}}
\caption{(Color online)
 \emph{Left:} the coexistence region (unshaded) 
in the ${\delta}_2$-$ {\delta}_0$ plane. 
Each point corresponds to a 
particular ratio $T_{m,0}/T_{c,0}$ 
and this ratio increases monotonically 
as $\delta_0$ grows 
at  fixed $\delta_2$.
\emph{Right:}
The value of $\Del C/\gamma T_c$ at the end point of the coexistence region, at $T_c
= T_{c,0} -0$ for $\delta_2/2\pi T_{c,0}=0.4, 0.28, 0.2$
 Thin
solid line is the BCS value $\Dc =1.43 \gamma$.  
Over some range of parameters, $\Del C/\gamma T_c$ at $T_{c,0}-0$ significantly
exceeds the BCS value. Diamonds represent the values $\Del C/\gamma T_c$ at $T_c=T_{c,0}-0$ for
the curves for $\delta_2/2\pi T_{c,0}=0.4$ and $0.28$ in Fig.~\ref{fig:2}}
\label{fig:3}
\end{figure}

We next consider $\Dc$ near the end point of the coexistence regime, 
 when 
$T_{m}\to T_{c,0} +0$,
$T_c\to T_{c,0}-0$ and $M_0$ is small.
In this case, we expand 
 $\alpha_\Delta$, $\eta$, and $\cF_0$ in 
terms of  
$M^2_0$,
use $\partial\alpha_\Delta(0,T)/\partial T=1/2T$, $\partial \alpha_\Delta/\partial M^2=2C_0$ and
$dM_0^2/dT = -(\partial\alpha_m/\partial T)/2B_0$
 and express 
\req{eq:heatjump} as
\begin{equation}
\label{eq:heatjumpGL}
\frac{\Del C}{T_c}=\frac{3\gamma}{2\pi^2 \eta_0}
\left(\frac{1}{2T}-\frac{\partial \alpha_m}{\partial T} \frac{C_0}{B_0}\right)_{T=T_c}^2,
\end{equation}
where 
 $\eta_0=A_0-C_0^2/B_0$ and the coefficients $A_0$, $B_0$ and $C_0$ are given by Eqs.~(\ref{eq:ABC}) with $M_0=0$. 
For $dM_0^2/dT$ we obtain from \req{eq:SCM0}
$dM_0^2/dT = -(\partial\alpha_m/\partial T)/2B_0$, \req{eq:SCM0}, with
\be
\frac{\partial \alpha_m}{\partial T} = \frac{1}{2T} - 2\pi \sum_{\vare_n>0}\left\langle
\frac{\delta_\varphi^2 \vare_n }{(\vare_n^2+\delta_\varphi^2)^2}\right\rangle .
\ee
In the absence of SDW order,  
$\Dc$ does not contain $(C_0/B_0)\partial \alpha_m/\partial T$,
 $\eta_0\to A_0$,
 and $\Dc= 1.43\gamma$.
Once $M_0$ is small but finite,  
$\Dc$  is determined by the interplay between   $(C_0/B_0)\partial \alpha_m/\partial T$ 
 and $C_0^2/B_0$ in $\eta_0$.
Both terms 
 originate from the fact that 
 SDW order 
 is suppressed 
 by SC order, $M^2 = M^2_0 - (C/B) \Delta^2$. 
As a result, $\Dc$  
is different from the BCS value 
 already for infinitesimally small SDW order and
exhibits discontinuity 
  upon  crossing the SDW border.
 In general, 
the change in $\eta_0$
 is more important
 because $C_0^2 = A_0B_0$
at perfect nesting, $\delta_{0,2}=0$,  and $\eta_0 =0$~\cite{fernandes,anton}.
Reduced value of $\eta$ leads to generally {\it larger} $\Dc$ at $T_c = T_{c,0} -0$ than $\Dc=1.43\gamma$ for transition from normal state to a pure SC state. As the coexistence region narrows, $\eta_0 \to 0$ and $\Dc$ grows.
Figure \ref{fig:3} shows $\Dc$ at $T_c = T_{c,0} -0$ for
 $\eta_0 >0$. In a wide range of parameters, $\Dc>1.43\gamma$.

\emph{Beyond mean-field}. In a mean-field description, $\Dc$ is discontinuous
at the tri-critical point $P$, Fig.~\ref{fig:2}, 
with $T_{m}=T_{c,0}$. 
The free energy $\cF_0$ and the specific heat jump depend on the finite
{\it square} of the SDW order, $M_0^2  \propto (T_{m}-T)$. Although above
$T_{m}$ the average $M_0=0$, 
one expects to replace $M_0^2$ by the finite {\it second} moment of SDW order due to 
Gaussian fluctuations,  $\langle M^2_0\rangle_{fluct}  \propto (T - T_{m})$. 
These fluctuations enhance $\Dc$ and transform the discontinuity in $\Dc$ into a
maximum, as shown in Fig. \ref{fig:1}.  As a result, $\Dc$ drops for deviations
from  $T_{c,0} = T_{m}$ 
both into the coexistence phase and away from the
SDW region.  Still, the decrease of $\Dc$ should be more rapid within the
SDW-ordered phase.  An enhancement of $\Dc$ by paramagnetic fluctuations was
earlier obtained in Ref.~\cite{kos}.

\emph{Conclusions.}
We demonstrated that the specific heat jump $\Dc$ 
across transition from SDW to the coexistence phase  significantly deviates from the BCS value. 
$\Dc$ has its maximum for  doping at the high-temperature end of the coexistence phase 
 and decreases for doping deviations in both directions. In the coexistence phase, $\Dc$ eventually becomes exponentially small as all low-energy states at low $T_c$ are already gapped out by the SDW order. At intermediate $T_c\lesssim T_{c,0}$, $\Dc$ scales 
 approximately as $T_c^2$.

This behavior is quite consistent with the observed doping evolution of $\Dc$ in   $Ba(Fe_{1-x}Ni_x)_2As_2$ and  $Ba(Fe_{1-x}Co_x)_2As_2$ \cite{budko_09}. In these materials $\Dc$ is peaked at the tri-critical point,
 which coincides with the optimal doping $x_{opt}$, and decreases for deviations from $x_{opt}$ in both directions, faster into the coexistence region. The 
  evolution of $\Dc$ there approximately follows $T_c^2$ behavior.

As doping increases and paramagnetic fluctuations disappear, $\Dc$ reduces to the BCS value.
Further reduction of $\Dc$ at larger dopings $x>x_{opt}$ is, most likely, a combination of several effects: 
(1) the enhancement of a non-magnetic interband impurity scattering~\cite{kogan,vvc_rho}; (2) 
stronger anisotropy of the gap on the electron FSs that increases $\eta$;  
(3) the reduction of $\gamma$ due to shrinking of the hole FSs.    

How strongly the value of $\Dc$ at $x_{opt}$ exceeds the BCS result is difficult to gauge because $\gamma$ has to be extracted  from the normal state
 $C(T)$ for which $\gamma T$ contribution is only a small portion 
 of the total specific heat.
 In  $Ba(Fe_{1-x}Co_x)_2As_2$,  $\Dc \sim 26~ mJ/(mol\cdot K^2)$ at $x_{opt}$, and  $\gamma\simeq 20~ mJ/(mol\cdot K^2)$~\cite{hardy_10}. 
In this case, the maximum of $\Dc$ 
 is not far from the BCS result, and the reduction of $\Dc$ into the coexistence state is mainly due to the gapped low-energy states by SDW order at low $T_c$.  
At the same time, a very strong $\Dc$ over a $100~ mJ/(mol\cdot K^2)$ near $x_{opt}$ in  $Ba_{1-x}K_xFe_2As_2$ is well above 
the BCS value~\cite{large}, even if  $\gamma$ is as large as 
 reported~\cite{large,hardy_10}  $50-60~mJ/(mol\cdot K^2)$.  This enhancement can well be due the peak in $\Dc$
 at the onset of the coexistence regime
or strong SDW fluctuations.

We thank S. Budko, R. Fernandes, I. Eremin, N. Ni, and J. Schmalian  for useful
discussions. M.G.V. is supported by NSF-DMR 0955500,  A.V.C. is supported by NSF-DMR 0906953, 
A.B.V. is in part supported by NSF-DMR 0954342.

\end{document}